\documentstyle[prl,aps,preprint]{revtex}

\begin{document}
%\twocolumn[

\preprint{RPI-94-N90}

\preprint{WM-94-106}

%\preprint{May 1994}

\title
{Leading-Log Effects in the Resonance Electroweak Form Factors}

\author{Carl E. Carlson$^{(a)}$ and Nimai C.
Mukhopadhyay$^{(b)}$}

\address{$^{(a)}$Physics Department,
College of William and Mary,
Williamsburg,  VA  23187}

\address{$^{(b)}$Department of Physics,
Rensselaer Polytechnic Institute,
Troy,  NY  12180-3590}

%\date{May 1994; this copy printed \today}
\date{\today}

\maketitle

%\vglue -0.3in

\begin{abstract}
\widetext
We study log corrections to inelastic scattering at high
Bjorken $x$ for $Q^2$ from 1 to 21 GeV$^2$. At issue is the
presence of log corrections, which can be absent if high $x$
scattering has damped gluon radiation. We find logarithmic
correction of the scaling curve extrapolated to low $Q^2$
improves the duality between it and the resonance plus
background data in the $\Delta$ region, indicating log
corrections exist in the data. However, at $W > 2$ GeV and high
$x$, the data shows a $(1-x)^3$ form. Log corrections in one
situation but not in another can be reconciled by a $W$- or
$Q^2$- dependent higher twist correction.
\end{abstract}
\vglue 0.2in

%]

\narrowtext

In this Letter, we investigate possible logarithmic
corrections to the inelastic structure function $\nu W_2$ at high
Bjorken $x$.

As a vehicle we use Bloom-Gilman (BG) duality, which is a
relationship
\cite{bg,dgp,cm90} between resonance physics and the physics of the deep
inelastic region. Bloom and Gilman \cite{bg}  observed that the ratio of
the area underneath a resonance bump in inelastic electron scattering to
that of the continuum beneath the bump was  generally constant with
increasing
$Q^2$ and that the smooth scaling curve seen at high $Q^2$ was an
accurate average over the resonance bumps seen at lower $Q^2$,  but the
same Bjorken $x$.   The first of these observations appears to be
untrue
\cite{cm93} for the
$\Delta(1232)$, although for other resonances it is well confirmed out
to high momentum transfer.  The second observation appears to be true in
general.  In particular, for the $\Delta(1232)$,  the background
seems to rise \cite{cm93}, as the resonance falls, so that the
average is constant relative to the scaling curve.  Theoretically,
one can understand in a perturbative QCD context~\cite{dgp,cm90} that the
$Q^2$ falloff of the resonance and of the scaling curve evaluated at the
$x$ value pertinent to the resonance are the same, at least as far as the
powers of $Q^2$  are concerned. The dependence on logarithms of
$Q^2$ has not yet been considered in this context.

	In this Letter, we first investigate the logarithmic corrections to
the resonance-continuum (or BG) duality discussed above. We compare
the resonance data \cite{s,bosted} to the scaling curve, for $Q^2$
from 1 to 21 GeV$^2$.  We correct the scaling curve using the
Altarelli-Parisi evolution equation \cite{ap}.  We find that the
corrections are sizeable and that they improve the duality between
the resonance data and the scaling curve.

We proceed by calculating the logarithmic corrections to the
predicted resonance form factors using common baryon distribution
amplitudes
\cite{cz}.  The logarithmic corrections that ensue do not track those
of the scaling curve.  We should not necessarily conclude that the BG
duality is violated by the logarithmic corrections,  having found
otherwise from the data.  Rather, this suggests a need for better
models of the baryon distribution amplitudes, or possibly that some
other effect is mimicking the log $Q^2$ behavior.

An issue is whether the logarithmic corrections to the quark
distributions coming from the Altarelli-Parisi equation are actually
present in the $x \rightarrow 1$ region relevant to low $W$ (hadronic
c.m. energy).  Brodsky {\it et al.}~\cite{b} have argued that the
gluon radiation that yields the splitting function, and the resultant
logarithmic corrections are absent in this region.  The criterion for
the absence of the gluon radiation is that $(1-x)Q^2 < \mu^2$, where
$\mu^2$ is some scale on the order of 1 GeV$^2$ but is not precisely
known.  We also note that $(1-x)Q^2 = x(W^2-m_N^2) \approx
(W^2-m_N^2)$ (for $x \rightarrow 1$),  so that the presence of log
corrections in the resonance region is an open question until $\mu^2$
is determined.

The measured values of $\nu W_2$ \cite{s} at high $x$ (above 0.7) and
for $W$ above the resonance region ($W > 2$ GeV) have a $(1-x)^3$
form.  No logarithmic corrections appear to be needed.  This is
consistent with the prediction of Brodsky {\it et al.}~\cite{b}, but
inconsistent with our observations in the $\Delta$ region. However,
the apparent absence of a logarithmic correction at high
$W$ and its importance in the resonance region can be reconciled by a
$W$- dependent higher twist correction.

{\it Logs in the continuum for $x \rightarrow 1$.}
We wish to see what effect
logarithmic corrections have on the parton distribution functions,
hence on the continuum scaling function. Since the resonance region
draws closer to  $x=1$ with increasing $Q^2$, we shall limit our
considerations for the continuum also to  $x \rightarrow 1$, and then
shall be able to quote some results in analytic form.

We start with the Altarelli-Parisi equation having unsuppressed
gluon radiation,

\begin{equation}
{{dq(x,t)} \over {dt}}={{\alpha _s(t)} \over {2\pi }}\int_0^1
{dydz\, \delta (x-yz)}q(y,t)P_{qq}(z).
\end{equation}
Here, $t= \ln(Q^2/\Lambda^2)$,  $\alpha_s(t) = 4\pi/\beta_1 t$,
$\beta_1 = 11-(2/3) n_f$,  (where $n_f$  is the number of fermion
flavors),  and $q(x,t)$ is a quark distribution function of a given
flavor. The gluon term is  omitted, as its contribution is subleading
in $(1-x)$.

	The splitting function is

\begin{equation}
P_{qq}(z)=C_F\left\{ {{{1+z^2} \over {(1-z)_+}}+{3 \over 2}\delta
(1-z)} \right\},
\end{equation}
where $C_F$ is $4/3$, and $(1-z)_+$ is defined by

\begin{equation}
\int_x^1 {dz{{f(z)} \over {(1-z)_+}}}=\int_x^1 {dz{{f(z)} \over
{1-z}}}-\int_0^1 {dz{{f(1)} \over {1-z}}}.
\end{equation}

We want to examine the evolution of a form like $q(x,t_0)=N_0(1-x)^b$,
where $b$ is a constant, and $t_0$ corresponds to some benchmark
$Q_0^2$. Hence we use the Ansatz:

\begin{equation}
q(x,t)=N(x,t)(1-x)^b,
\end{equation}
in the Altarelli-Parisi equation.  Systematically throwing away terms
of higher order in  $(1-x)$,we get
\begin{equation}
q(x,t)=N_0(1-x)^{b+{{4C_F(\ln \ln Q^2)} / {\beta _1}}},
\end{equation}
where
\begin{equation}
\ln \ln Q^2\equiv \ln \left( {{\ln Q^2 / \Lambda ^2}
\over {\ln Q_0^2 / \Lambda ^2}} \right)   \equiv T(Q).
\end{equation}
To see the size of the logarithmic correction, we examine the values
of $x$ corresponding to the peak of the $\Delta$ resonance region,
\begin{equation}
{1 \over x}=1+{{m_\Delta ^2 - m_N^2} \over {Q^2}}.
\end{equation}
In Table \ref{table1}, we show the size of the correction factor to
the high-$x$ continuum in the  $\Delta$-region using $Q^2$
of 4 GeV$^2$ as a benchmark.  The logs are important for the resonance
excitation even though the  correction to the exponent is fairly
mild.  For example,  the $(1-x)^3$ we use for $F_2$ at $Q^2$ of 4
GeV$^2$ is modified by radiative corrections to about $(1-x)^{3.16}$
at $Q^2
\approx 20$ GeV$^2$.  However,  the value of $(1-x)$ is very small
for the $\Delta$ at the latter $Q^2$ leading to a change in $F_2$ by a
factor $(0.029)^{0.16} \approx 0.57$.

	We close this section with two observations. One concerns a
parameterization of the quark distribution by Morfin and Tung
\cite{mt}. It is of the form

\begin{equation}
q(x,t)\propto (1-x)^{C_0+C_1T(Q)+C_2T^2(Q)},
\end{equation}
for $x \rightarrow 1$.  Here the coefficient $C_1$ is just what is here
calculated to be $(4C_F/\beta_1)$. For their DIS-scheme fits,  we
note,  they get $C_1 =0.53-0.54$, whereas

\begin{equation}
{4C_F \over \beta_1}=0.59\;,\;0.64,
\end{equation}
for the three or four flavors respectively. Our second remark concerns
the comparison of $F_2=\nu W_2$ data in the resonance region to the
continuum scaling curve. This is shown in Fig. \ref{fig1}, with and
without the log correction.  In each case, the ratio  $R=I/S$ is
plotted versus
$Q^2$. Here we have defined

\begin{eqnarray}
  I&=&\int\limits_{\Delta \xi } {d\xi }\ F_2(\xi,Q^2) , \nonumber\\
  S&=&\int\limits_{\Delta \xi } d\xi \  F_2^{scaling}(\xi,Q^2),
\end{eqnarray}
where $\xi$ is the Nachtmann variable ($x$ corrected \cite{cm93,n} for
the target mass effects),  and $\Delta \xi$ is a bite covering the
chosen resonance region (here, the $\Delta$). Although the error bars
are large at high
$Q^2$,  one observes that the logarithmic corrections improve the
constancy of the ratio $R$ at high $Q^2$.  Hence the inclusion of the
logarithmic effects helps to make the duality idea,  the low-$Q^2$
structure function for a given $W$ should average to the scaling curve,
appear to work better.

{\it Log $Q^2$ effects on the resonance to continuum ratio.}
The resonance contribution to the inelastic structure function is

\begin{equation}
F_2^R=G_+^2(Q^2){{m_N^2\Gamma _R /  \left( 2\pi {m_R} \right)} \over
{(W-m_R)^2+\Gamma _R^2 / 4}},
\end{equation}
assuming a simple Breit-Wigner form , and dropping  the sub-leading
helicity  form factors $G_0$ and $G_-$. The form factor is

\begin{equation}
G_+(Q^2)=g_+{{\alpha _S^2(Q^2)} \over {Q^3}}\sum
{(E_{ij}N_i^PN_j^R)(\ln Q^2)^{-\gamma _i^P-\gamma _j^R}},
\end{equation}
where $N_P^i$ and $N_R^j$ are coefficients from the distribution
amplitudes of the proton and resonance respectively. The latter is,
for  example, given by

\begin{equation}
\Phi ^R(x,Q^2)=x_1x_2x_3\sum\limits_i {N_i^R}\tilde \Phi _i(x)(\ln
Q^2)^{-\gamma _i}.
\end{equation}
The $\Phi^i$ are Appel polynomials and the anomalous dimensions
$\gamma^i$ are known and positive.
	We take the form of the continuum for $x\rightarrow 1$ as
\begin{equation}
F_2^{scaling}(x,Q^2)={const} (1-x)^{b+C_1 T}.
\end{equation}
We compare the resonance and continuum contribution to the ratio of
integrals $R=I_i/S_i$. First,

\begin{equation}
S_i=\int\limits_{\Delta x_i} {dxF_2^{scaling}}\approx {{const} \over
{Q^8}}(\ln Q^2)^{-C_1\ln Q^2},
\end{equation}
for $b=3$. The other integral is

\begin{eqnarray}
I_i &=& \int\limits_{\Delta x_i} {dxF_2^R}\approx {{const} \over
{Q^2}}\left| {G_+} \right|^2 \nonumber \\
&\approx& {{const} \over {Q^8}}\left|
{\sum\limits_{ij} {(E_{ij}N_i^PN_j^R)(\ln Q^2)^{-2-\gamma _i^P-\gamma
_j^R}}} \right|^2,
\end{eqnarray}
where we have recalled $\alpha_s  \sim  1/\ln Q^2$.

	It is clear that $I_i$ and $S_i$ have the same power law falloff (for
$F_2$ scaling $\sim (1-x)^3$  for $x \rightarrow  1$).  The $\ln Q^2$
dependences can be approximately the same only under special
circumstances and a limited range of the $Q^2$.  Let us consider only a
small range of
$\ln Q^2$ ($\equiv {{\ln (Q^2 / \Lambda ^2)} {\big /} \,{\ln (Q_0^2 /
\Lambda ^2)}}$) and expand around $\ln Q^2 = \ln Q_0^2$,  so that $\ln
Q^2 = 1+\epsilon$.  Then equating the O(1) and O($\epsilon$) terms of the
expansion leads to the relation

\begin{equation}
{1 \over 2}C_1={{\sum\limits_{i\,j} {(E_{ij}N_i^PN_j^R)(2+\gamma
_i^P+\gamma _j^R)}} \over {\sum\limits_{i\,j} {(E_{ij}N_i^PN_j^R)}}}.
\label{17}
\end{equation}

Since $C_1/2$ is about 1/4 and the $\gamma_i$ are positive,  only
exceptional choices of the amplitudes $N_i^P$ and/or $N_j^R$  can
fulfill the above equation.  We know no cases of practical that
satisfy Eqn.~\ref{17}.  For instance, it is not fulfilled for the cases
\cite{cz} of the Chernyak-Zhitnitsky or the King-Sachrajda wave function
for the proton, and analogous wave functions for the S$_{11}$(1535) or
the
$\Delta(1232)$.  It happens that every significant
$E_{ij} N_i^P N_j^{S_{11}}$ is positive, violating Eq.(17).  Thus, in
general, the BG duality in the form of the constancy of the resonance
peak to scaling curve ratio must be  logarithmically  violated at high
$Q^2$. The resonance will fall  faster  than the background. The main
reason for this is the ($\alpha_s(Q^2))^2$ factor in the exclusive  state
form factor, absent for the inclusive process. This quantity falls  by a
factor of nearly two [1.85 for $\Lambda_{QCD}=200$ MeV],  as
$Q^2$  changes from 4 to 21 GeV$^2$.  That $Q^3 G_+(Q^2)$ is nearly
constant in this range of $Q^2$ is interesting and not understood at
this level.

{\it The phenomenological parton distribution functions.}
For our
purposes, we need the parton distribution functions for $x\rightarrow 1$.
However,the existing parameterizations
\cite{mt,ehq} are fit to data at lower $x$, and thus are not designed to
be outside specified ranges of $x$. For example,  both Morfin and Tung
\cite{mt} and Botts {\it et al.}\ (the CTEQ collaboration)~\cite{cteq}
state their fits to be valid for
$x < 0.75$; similar restrictions apply to other
parameterizations. Hence, some diffidence is required in extrapolating
these functions toward
$x=1$, and one should not be surprised by disagreements among various
parameterizations, and between any of them and the data, as $x
\rightarrow  1$.

Fig. \ref{fig2} shows some high $x$, non-resonance region data
\cite{s}.   The parameterizations of Morfin and Tung~\cite{mt} and
CTEQ~\cite{cteq} are also shown.  They fall too rapidly in this region
and are below the data by a factor of roughly two at the highest $x$ data
point.  The naive, uncorrected $(1-x)^3$ curve matches the data
better.   We should note that the logarithmic corrections will not give
as dramatic an effect here as in the resonance region.  For the example
of the points in Fig. \ref{fig2},  where $Q^2$ is about $20$
GeV$^2$,  a radiative correction factor of $(1-x)^{0.16}$ falls
from 0.83 to 0.74 as we go from the lefthand data point to the right
hand data point.  That is a change of barely over 10\%,  although
including it worsens the agreement with the data.

Here we recall the Brodsky {\it et al.} \cite{b} argument that the
logarithms are healed (absent) for kinematics where
$(1-x)Q^2$ is small.  The high $x$ non-resonance region data
does favor the Brodsky {\it et al.} suggestion.  However,  we
already have seen the importance of the the QCD radiative
corrections in the $\Delta$ region.  The two seemingly disparate
observations could be reconciled by allowing a $W$-dependent higher
twist correction,  so that we have in the high $x$ region
\begin{equation}
F_2 \propto (1-x)^{3 + 4C_F T(Q)/\beta_1}\times
            \left(1+C_2 {m_N^2 \over W^2}\right),
\end{equation}
where $C_2 = 1.7$ gives the dashed curve in Fig. \ref{fig2}.

{\it Concluding remarks.}
We have studied here the leading log QCD radiative corrections at high
$Q^2$ in and near the resonance region.  The region of the
$\Delta(1232)$ resonance, in which the resonance bump falls faster
than the underlying background, is of special interest.  We have
found the logarithmic corrections to be important for $Q^2 > 4$
GeV$^2$.  The agreement between the $F_2$ data in the resonance
region,  smoothed over the resonance width,  and the scaling curve is
much improved by the logarithmic corrections.  This indicates that
the gluonic radiative corrections are important even in the
resonance region.

We have also considered the effect of the log corrections to the
baryon form factors directly.  In general,  these effects are
dependent on the specific baryon wave functions.  Log corrections
to form factors,  for commonly used baryon distribution amplitudes
\cite{cz}, disagree with those to the scaling curve.  Accepting the
common distribution amplitudes means that the resonance-background
duality is violated logarithmically,  contradicting the observations
summarized in Fig. \ref{fig1}.  This suggests the need for better
model baryon distribution amplitudes.

The measured $F_2 = \nu W_2$ at high $x$ and $W > 2$ GeV
is nicely fit by a plain $(1-x)^3$ form.  This agrees with the
expectation of Brodsky {\it et al.}~\cite{b}, although $(1-x)Q^2
\approx 3$ GeV$^2$ for this region, which is large for the absence
of gluonic radiation.  The apparent absence of a log correction here
is also surprising in the light of its apparent presence in the
resonance region,  but these can be reconciled by a higher twist
correction of the form $(1+const\ m_N^2/W^2)$.

One could entertain an alternative explanation of our observations:
Log corrections coming from gluonic radiation are absent
{\it everywhere} in the high $x$ regions we have studied, and a $Q^2$
dependent higher twist correction is giving the effect we have
observed for the $\Delta$ region.  In either case, higher twist
corrections are indicated, with different kinematic dependence.

Our conclusions invite more precise and complete experimental
tests.  In particular,  the hypothesis on evolution healing
\cite{b} can be tested against the full leading log corrected
structure function ameliorated by higher twist corrections
discussed here,  by measuring that structure function over a range of
high $x$ at fixed values of $W$ \cite{footnote} in one instance and
at fixed $Q^2$ in another.

\bigskip
{\underline {Acknowledgments}}.  We thank S. Brodsky and P.
Stoler for helpful comments.  The research of CEC is supported in part
by the NSF (Grant PHY-9306141),  and that of NCM by the U.S.
Department of Energy (Grant DE-FG02-88ER40448.A006).

\begin{table}
\caption{Numerical values for the correction to $(1-x)^b$ for
the $\Delta(1232)$ excitation kinematics.  The function $T(Q)$
is defined in the text.  We used $\Lambda = 150$ MeV.}
\vskip 0.0in

\begin{center}
\begin{tabular}{ccc}

$Q^2$ GeV$^2$ & $x$ & $(1-x)^{4C_F T(Q)/\beta_1}$\\  \hline

  4   &    0.863  & 1.000     \\
  6   &    0.904  & 0.901     \\
  8   &    0.926  & 0.824  \\
 10   &    0.940  & 0.761  \\
 12   &    0.950  & 0.711  \\
 17   &    0.964  & 0.615  \\
 21   &    0.971  & 0.560
\end{tabular}

\end{center}

\label{table1}
\end{table}

\begin{figure}

\vglue 3.5in  %ratio
\hskip 0.2in {\special{picture fig1 scaled 1000}} \hfil
\vglue 0.3in
\caption{The effect of logarithmic corrections upon the duality
ratio $R$,  defined in the text.  The heavy circles with
uncertainty bars indicate $R$ with log corrections made
for the scaling curve; the open circles indicate where central
values lie when no log corrections are made (percentage
uncertainties are the same).  The ratios with corrections lie
more closely on a horizontal line, as as predicted by
perturbative QCD, drawn here with arbitrary ordinate.}
  \label{fig1}
\end{figure}

\newpage

\begin{figure}

\vglue 3.5in  %high x F_2
\hskip 0.2in {\special{picture fig2 scaled 1000}} \hfil
\vglue 0.3in

\caption{Measured $F_2 = \nu W_2$~\protect\cite{s} at
high $x$ above the resonance region ($W > 2$ GeV),  represented by
triangles.  The solid line is
$(1-x)^3$,  the tight dashed line is the parameterization DIS of
Morfin and Tung~\protect\cite{mt},  the dash-triple-dotted line (close to
the Morfin-Tung line) is from the CTEQ1L
distribution~\protect\cite{cteq}, and the loose dashed line (close to the
solid line) is a result including both logarithmic corrections and a $W$
dependent higher twist correction,  as described in the text.  Values of
$Q^2$ range from 16 to 19 GeV$^2$ for the data in this figure, and
values of $W$ range from 2.8 to 2.0 GeV.  Uncertainties in the data are
about $\pm$10\%,  or about the size of the triangles.}
  \label{fig2}
\end{figure}


\begin{thebibliography}{99}
\bibitem{footnote} For example,  fixing $W$ at 2 GeV$^2$ and letting
$x$ vary from 0.85 to 0.75 changes the radiative correction factor,
roughly $(1-x)^{0.6 T(Q)}$,  by about 20\%.  The corresponding $Q^2$
are 18 and 9.4 GeV$^2$, respectively.
\end{thebibliography}
\end{document}